\documentclass[journal,twoside,web]{ieeecolor}
\usepackage{jsen}
\usepackage{cite}
\usepackage{amsmath,amssymb,amsfonts}
\usepackage{algorithmic}
\usepackage{graphicx}
\usepackage{textcomp}
\usepackage{caption} 
\usepackage{wrapfig}
\def\BibTeX{{\rm B\kern-.05em{\sc i\kern-.025em b}\kern-.08em
    T\kern-.1667em\lower.7ex\hbox{E}\kern-.125emX}}
\definecolor{abstractbg}{rgb}{0.89804,0.94510,0.83137}
\begin{document}
\title{Lower Limb Movements Recognition Based on Feature Recursive Elimination and Backpropagation Neural Network}
\author{Yongkai Ma, Shili Liang, Zekun Chen
\thanks{We are very grateful to Northeast Normal University for providing the experimental platform. This research is supported by the Jilin Provincial Science and Technology Department [grant number YDZJ202201ZYTS506].(Corresponding author: Shili Liang.)}
\thanks{Yongkai Ma and  Zekun Chen are with the School of Physics, Northeast Normal University, Changchun 130024, China (e-mail: mayk448@nenu.edu.cn) }
\thanks{ Shili Liang is also with the School of Physics, Northeast Normal University, Changchun 130024, China (e-mail: lsl@nenu.edu.cn)}}

\IEEEtitleabstractindextext{%
\fcolorbox{abstractbg}{abstractbg}{%
\begin{minipage}{\textwidth}%
\begin{wrapfigure}[11]{r}{3in}%
\includegraphics[width=3in]{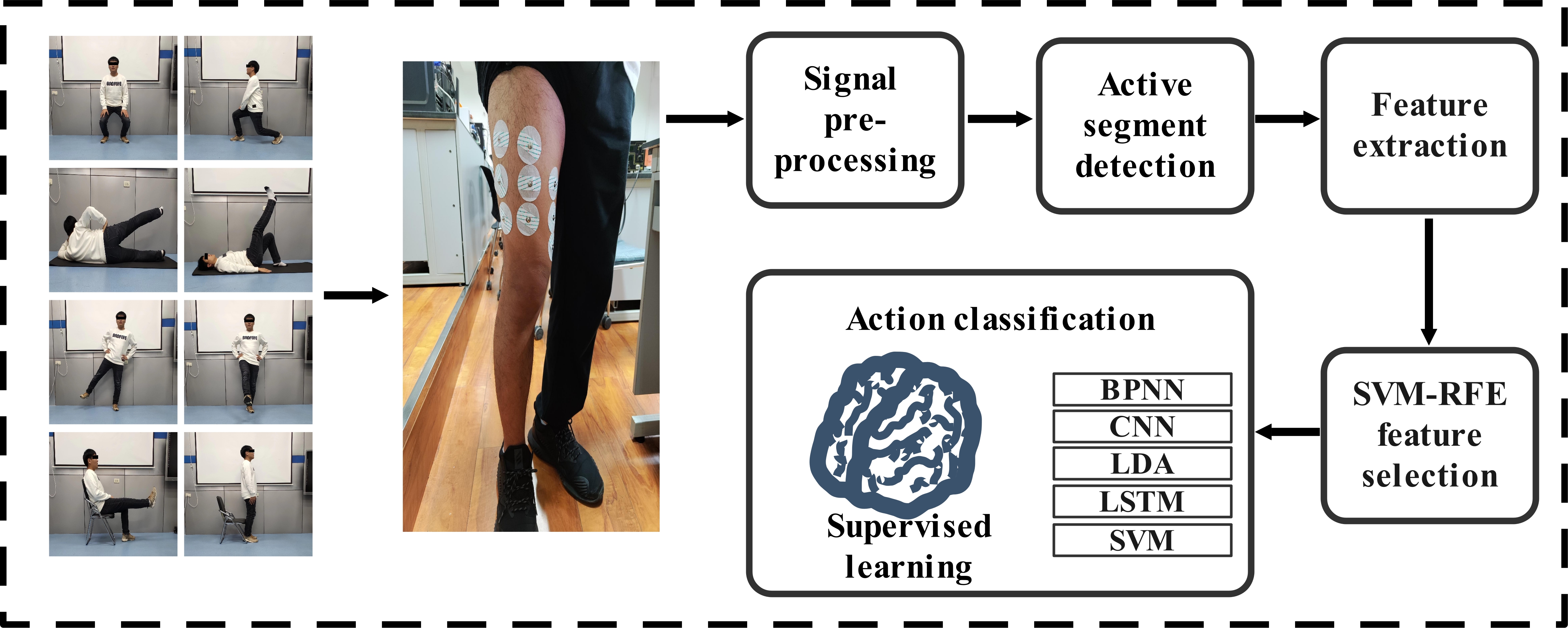}%
\end{wrapfigure}%
\begin{abstract}
Surface electromyographic (sEMG) signal serve as a signal source commonly used for lower limb movement recognition, reflecting the intent of human movement. However, it has been a challenge to improve the movements recognition rate while using fewer features in this area of research area. In this paper, a method for lower limb movements recognition based on recursive feature elimination and backpropagation neural network of support vector machine is proposed. First, the sEMG signal of five subjects performing eight different lower limb movements was recorded using a BIOPAC collector. The optimal feature subset consists of 25 feature vectors, determined using a Recursive Feature Elimination based on Support Vector Machine (SVM-RFE). Finally, this study used five supervised classification algorithms to recognize these eight different lower limb movements. The results of the experimental study show that the combination of the BPNN classifier and the SVM-RFE feature selection algorithm is able to achieve an excellent action recognition accuracy of 95\%, which provides sufficient support for the feasibility of this approach.
\end{abstract}

\end{minipage}}}

\maketitle

\section{Introduction}
\label{sec:introduction}
\IEEEPARstart{S}{urface} electromyography (sEMG) signal is a weak bioelectric signal produced by muscle fiber units during muscle contraction \cite{staude1999objective}. The sEMG signal correlates with the level of muscle contraction in the human body. The sEMG signal is easy to collect and can be recorded in a non-invasive way to record muscle activity \cite{thongpanja2012feasibility}, using disposable surface electrodes attached to the skin. The sEMG signal has received widespread attention in disease diagnosis, rehabilitation training, exoskeleton robotics, human-computer interaction, and bionic prosthetics due to its advantages of easy collection, non-invasive, and natural \cite{zwarts2000recent,barry1990acoustic,singh2012trends,qi2019intelligent,gauthaam2011emg}.

The intention behind human movements is directly reflected in the sEMG signal, which is often used as input for human-computer interaction or pattern classifiers \cite{artemiadis2012emg,khezri2007real}.The process of limb movement recognition using the sEMG signal is generally divided into two processes, i.e., feature extraction and model classification. Through feature extraction, we are able to extract useful information from the sEMG signal and filter out unwanted parts and interferences \cite{phinyomark2012feature}. When processing the sEMG signal, we often use time domain analysis, frequency domain analysis, wavelet transforms, Fourier transforms, and autoregressive model for feature extraction. In the study of EMG signal classification, the appropriate feature vectors must be carefully selected to achieve better classification performance. Too many redundant features can lead to the degradation of classifier performance, increase the computational burden, and cause overfitting problems \cite{kwak2002input}. In the area of feature selection, extensive research has been conducted on various algorithms, including variance selection, correlation coefficient analysis, mutual information computation, recursive feature elimination, and tree-based feature selection \cite{guyon2002gene,saranya2023resistive,yan2008application,yin2014operator,zhou2021feature}.

In recent years, the Recursive Feature Elimination based on Support Vector Machine (SVM-RFE) has been widely used in the sEMG signal and EEG signal feature selection with good classification results. For example, in \cite{lee2022gender}, the authors proposed the identification of four gait features by SVM-RFE and Feature Recursive Elimination based on Random Forest (RF-RFE). The results show that RF-RFE achieved a recognition rate of 98.89\%, while SVM-RFE reached 99.11\%. Notably, SVM-RFE showed superior performance. In their work \cite{tosin2017semg}, the authors proposed using SVM-RFE to determine the optimal feature subset, which improved the recognition rate of wrist/finger movements to over 90\%. In their study \cite{tosin2020statistical}, the authors conducted a comparison of three feature selection methods, namely SVM-RFE, Monte Carlo, and Singular Value Decomposition Entropy Ranking, for movements recognition using a Regularized Extreme Learning Machine (RELM). The obtained average accuracies were 84.9\%, 84.0\%, and 83.9\%, respectively. These results highlight that the SVM-RFE feature selection method achieved the highest recognition rate. In their work \cite{dadebayev2021machine}, the authors used recursive feature elimination to identify the optimal subset of features and applied SVM to classify human movement states using EEG signals.

When it comes to action classification of the sEMG signal, several classifiers have been developed. A very comprehensive description of the application of SVM to the sEMG signal is presented in \cite{toledo2019support}. In \cite{bukhari2020study}, the authors used a k-nearest neighbor (KNN) classifier to classify different levels of muscle contraction and showed low classification accuracy in fatigue.  In \cite{song2019design}, the authors used a Decision Tree (DT) to classify 12 gestures with 91\% accuracy.In \cite{zhang2013adaptation}, the authors used a Linear Discriminant Analysis (LDA) classifier to classify different motion patterns based on the sEMG signal. In \cite{chaobankoh2022lower}, the authors used the spectrograms of three-foot movements as input to a Convolutional Neural Network (CNN), which ultimately classified the three-foot movements. In \cite{wang2020semg}, the authors used the Long Short-Term Memory (LSTM) network to classify and estimate six daily hand movements.In \cite{ahsan2011neural}, the authors used a Backpropagation Neural Network (BPNN) to recognize multiple hand movements.  These classifiers have individual strengths and suitability in different application contexts and serve as valuable tools for the sEMG signal classification and recognition. 

The major contributions of this work include:

First, we created a database of the sEMG signal associated with lower limb movements using the BIOPAC acquisition device, which included sEMG signal data for eight lower limb movements from five subjects.

Second, by using the multi-threshold activity segment detection method proposed in this paper, we are able to detect the start and end points of the sEMG signal corresponding to each action, which facilitates the subsequent feature extraction.

Finally, by selecting the best subset of features through SVM-RFE, we improved the accuracy of lower limb movements classification, which provided critical support for this study.

In this study, we extracted the best subset of features using the SVM-RFE algorithm and fed this best subset of features into five different supervised classifiers for lower limb movement classification.The experimental results show that the combination of the SVM-RFE feature selection method and BPNN classifier exhibits excellent recognition performance and successfully improves the accurate recognition rate of eight lower limb movements to 95\%.

The paper is organized as follows. Section II describes the experimental program in detail. In Section III, we propose a multi-threshold active segment detection method, followed by the use of SVM-RFE to select the optimal subset of features, and then five supervised learning classification models are introduced. Section IV shows and analyzes the classification results of the five supervised classifiers for lower limb movements. Section V summarizes and discusses the paper.

\begin{figure}[htbp]
	\centering
	\includegraphics[scale=0.35]{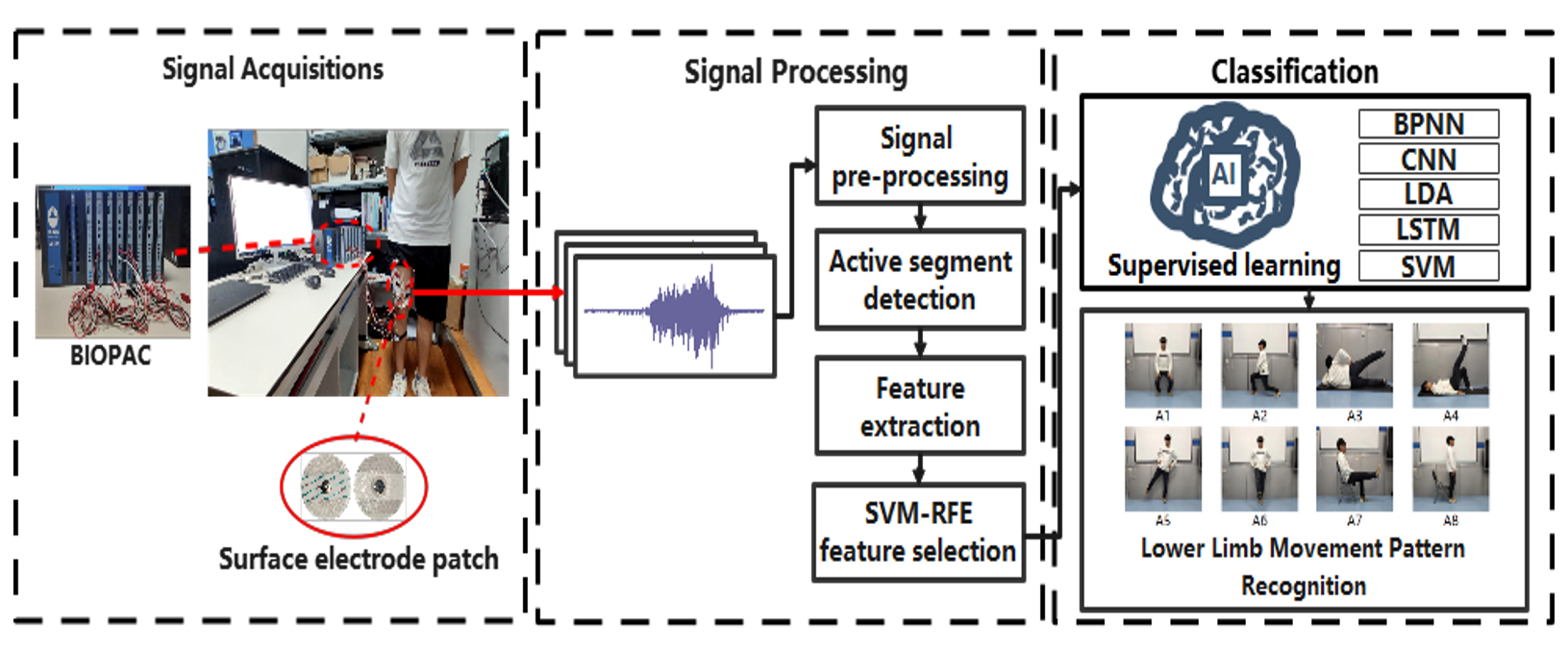}
	\caption{Flowchart of the overall study}
	\label{fig-1}
\end{figure}

The overall research process is shown in Fig \ref{fig-1}.

\section{Materials and Methods}
\subsection{Experimental program}
\subsubsection{ Muscle selection}
\begin{figure}[htbp]
	\centering
	\includegraphics[scale=0.58]{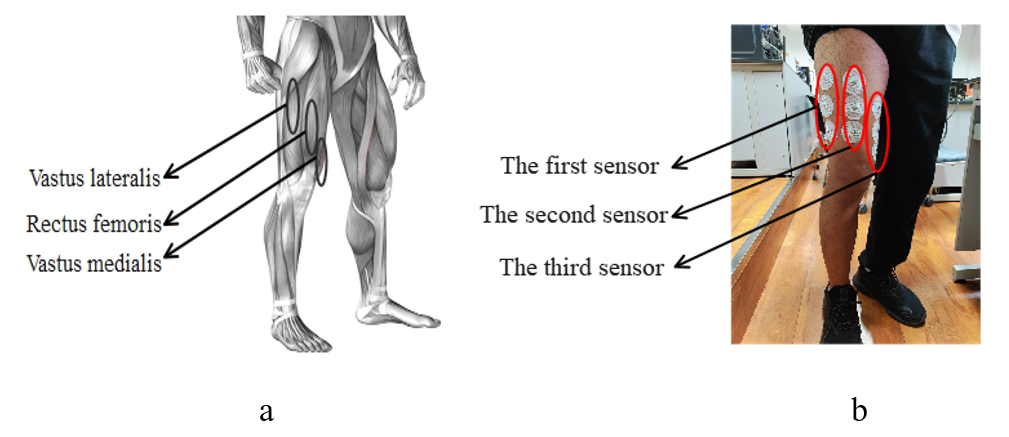}
	\caption{Muscle and sensor wearing position. (a) Distribution of muscle groups in the lower foreleg; (b) Sensor positions for each subject}
	\label{fig-2}
\end{figure}

Before applying the sensors, the skin surface of the recording area was cleaned with alcohol wipes to ensure good contact with the muscles of the lower limbs, to reduce the noise generated by the offset of the electrode pads during signal acquisition, and to improve the quality of the signal acquisition. Disposable AgCl electrode patches were used in the experiments, which can reduce interference from nearby muscles due to their small area \cite{mesin2009effect}. We placed three sensors in the rectus femoris, lateral femoris, and medial femoris muscles of the participant's right leg, and the muscle locations and actual electrode patch wearing positions are shown in Fig \ref{fig-2}.

\subsubsection{ Acquisition process}

In this study, five subjects, both male and female, ranging in age from 23 to 27 years were recruited. 

All subjects received special instruction and training before data collection to ensure that the lower extremity movements were completed at a smooth pace. Before performing the lower limb movements, the subjects' thigh muscles remained in a relaxed state and returned to a relaxed state after the lower limb movements were completed. It should also be noted that all subjects read and signed an informed consent form before participating in the experiment and had no known physical problems.

\begin{figure}[htbp]
	\centering
	\includegraphics[scale=0.6]{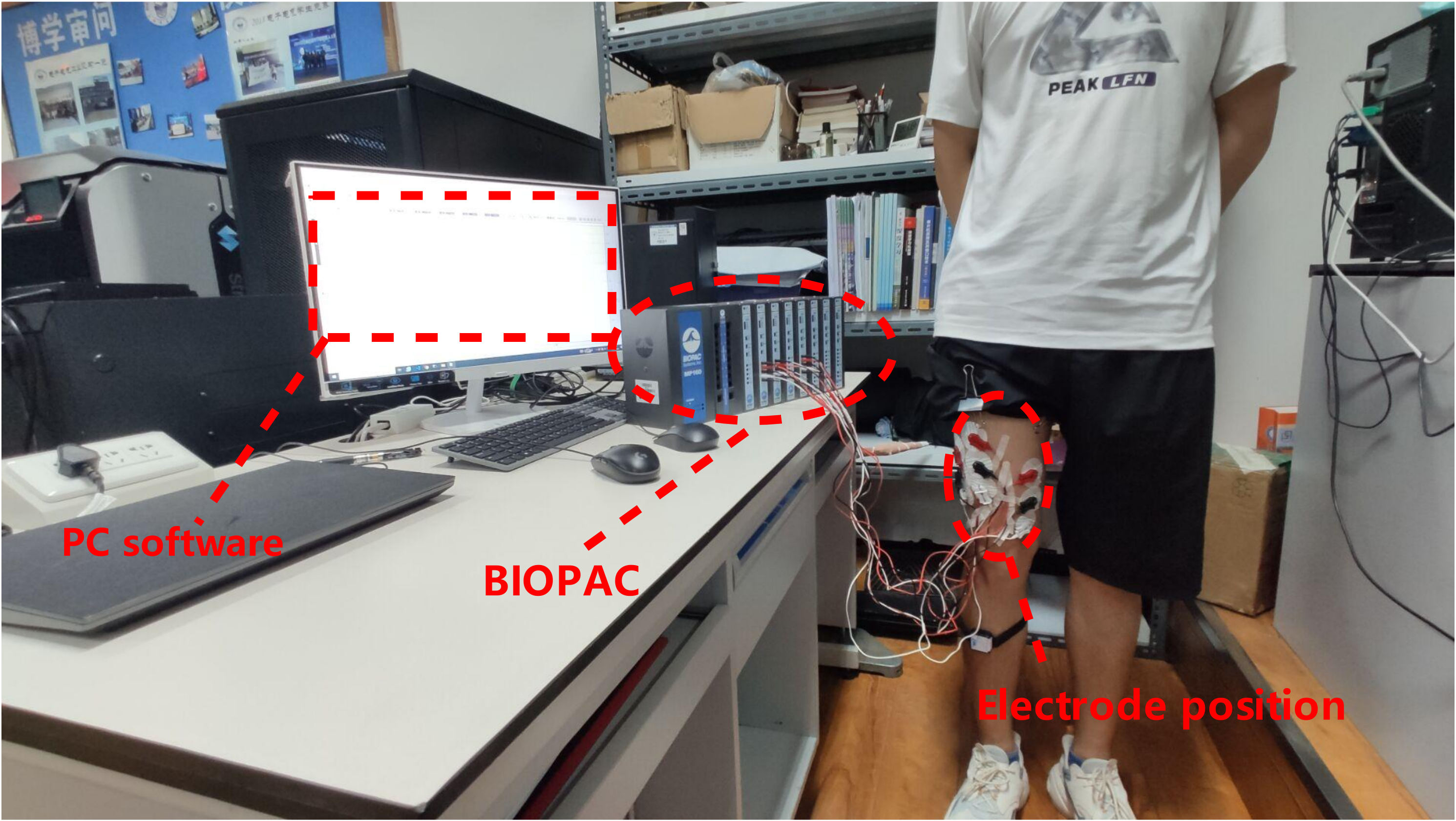}
	\caption{Actual acquisition process}
	\label{fig-3}
\end{figure}

In this study, we used the BIOPAC acquisition device to collect the data, and the sampling rate of the sEMG signal was 2000 Hz. The subjects were able to observe the changes in the sEMG signal in real-time through the upper computer interface, and the actual acquisition process is shown in Fig \ref{fig-3}.

\begin{figure}[htbp]
	\centering
	\includegraphics[scale=0.8]{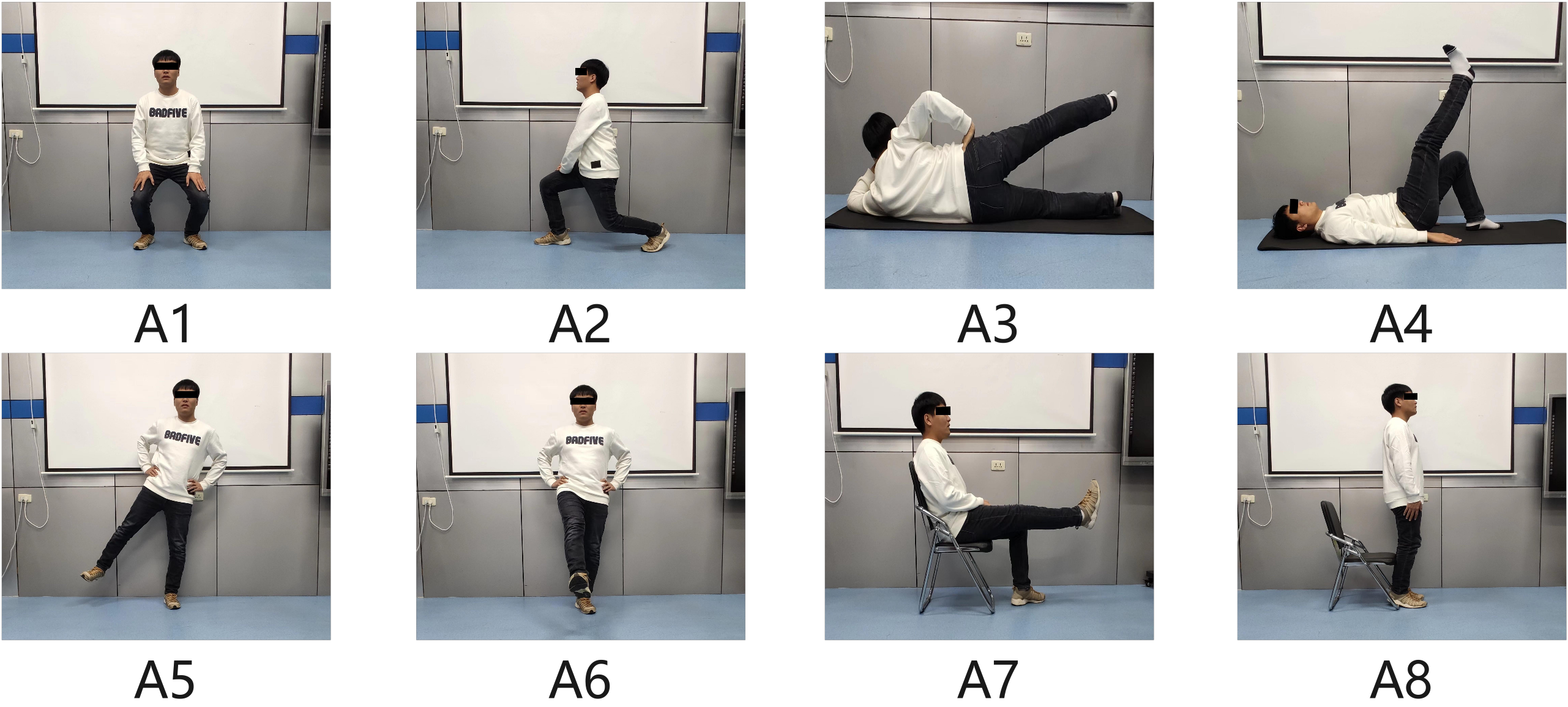}
	\caption{8 lower limb movements}
	\label{fig-4}
\end{figure}

In this experiment, eight common lower limb movements were investigated, as shown in Fig \ref{fig-4}, each movement lasted for three seconds, and the eight movements were as follows: the first action (A1): squatting, the second action (A2): bowing, the third action (A3): lying lateral kicking, the fourth action (A4): lying straight kicking, the fifth action (A5): standing lateral kicking, the sixth action (A6): standing straight kicking, the seventh action (A7): sitting kicking, and the eighth action (A8): sitting standing sitting. 

In this experiment, each movement had to be repeated 60 times by all subjects. A total of 480 samples per subject were collected from 8 different lower limb movements. These 480 samples were randomly divided, of which 384 samples (80\%) were randomly selected as the training set, while the remaining 96 samples (20\%) were also randomly divided as the test set. Table \ref{tab1} provides detailed information about the sample distribution.

\begin{table}[h!]
	\centering
	\scriptsize
	\caption{Distribution of samples}
	\begin{tabular}{p{1.3cm} *{10}{c}}
		\hline
		\textbf{Class} & \textbf{A1} & \textbf{A2} & \textbf{A3} & \textbf{A4} & \textbf{A5} & \textbf{A6} & \textbf{A7} & \textbf{A8} &\textbf{Total} \\ 
		\hline
		\textbf{Train set}           & 48          & 48          & 48          & 48          & 48          & 48          & 48          & 48          & 384            \\
		\textbf{Test set}            & 12          & 12          & 12          & 12          & 12          & 12          & 12          & 12          & 96             \\
		\textbf{Total}                  & 60          & 60          & 60          & 60          & 60          & 60          & 60          & 60          & 480            \\ \hline
	\end{tabular}
	\label{tab1}
\end{table}

\section{Signal Preprocessing}
\subsection{ Signal Filtering}
\begin{figure}[htbp]
	\centering
	\includegraphics[scale=0.45]{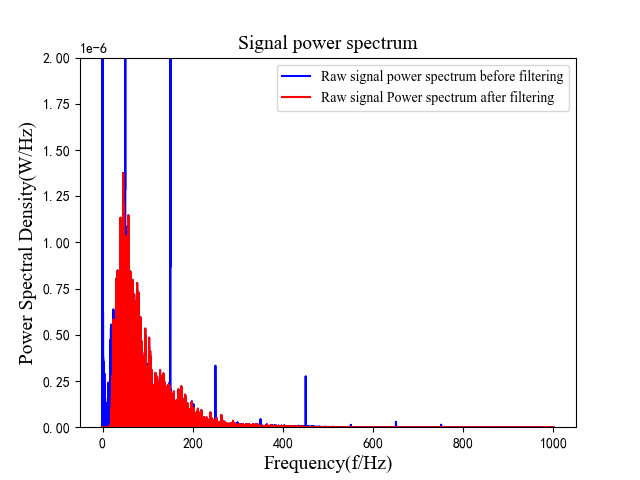}
	\caption{Comparison of the sEMG signal power spectrum before and after filtering}
	\label{fig-5}
\end{figure}

The raw sEMG signal is a mixture of the noise signal and the sEMG signal \cite{phinyomark2009novel}. The noise signal includes the intrinsic noise caused by industrial interference and the artifact noise caused by human movements. To remove the inherent noise and its harmonic components, we used third-order Butterworth band-stop filters set at frequencies of 49-51Hz, 99-101Hz, 149-151Hz, 199-201Hz, 249-251Hz, 299-301Hz, 349-351Hz, 399-401Hz, and 449-451Hz, respectively. Movements artifacts occur mainly appear below 20 Hz \cite{conforto1999optimal}. The main energy range of the sEMG signal is between 20 Hz and 500 Hz, so in this study, to remove the components not relevant to the study, we used an 8th-order zero-phase Butterworth filter with a bandwidth between 20 Hz and 500 Hz \cite{restrepo2017improving,guo2015nonlinear,cid2018normalization}. Fig \ref{fig-5} shows the power spectrum of the signal before and after filtering, before filtering in blue and after filtering in red.
\subsection{Active segment detection}
Activity segment detection is a key step in the processing of the sEMG signal and by effectively detecting the start and end positions of each activity, the amount of data processing for subsequent feature extraction can be reduced \cite{chen2013pattern}.Compared with traditional algorithms using the short-term energy method to detect active segments, this paper proposes to use the multi-threshold short-term energy method for active segment detection, which can effectively detect active segments, and its method includes the following three steps:

Step 1: Assuming n channels, if $x_{kij}$ represents the i-th sampling point of the k-th active segment in the j-th channel, there are a total of 32 sampling points in the following 16 milliseconds, with a sampling rate of f = 2000 Hz. Therefore, using a window width of 32, the average short-term energy calculated in this 16 milliseconds period can be expressed as follows:
\begin{equation}
	{{\bar{E}}_{k}}=\frac{1}{32n}\sum\limits_{i=0}^{31}{\sum\limits_{j=1}^{n}{x_{kij}^{2}}}
	\label{eq:1}
\end{equation}
Step 2: Select appropriate thresholds TH1 and TH2 according to the experimental requirements and signal characteristics, TH1 indicates the start threshold and TH2 indicates the stop threshold. If the condition ${{\bar{E}}_{k}} > TH1$ is satisfied, the action starts and marks the start point of the action as $x_{s}$, and if  the condition ${{\bar{E}}_{k}} < TH2$ is satisfied, the action ends and marks the endpoint of the action as $x_{e} (e > s)$.

\begin{figure}[htbp]
	\centering
	\includegraphics[scale=0.5]{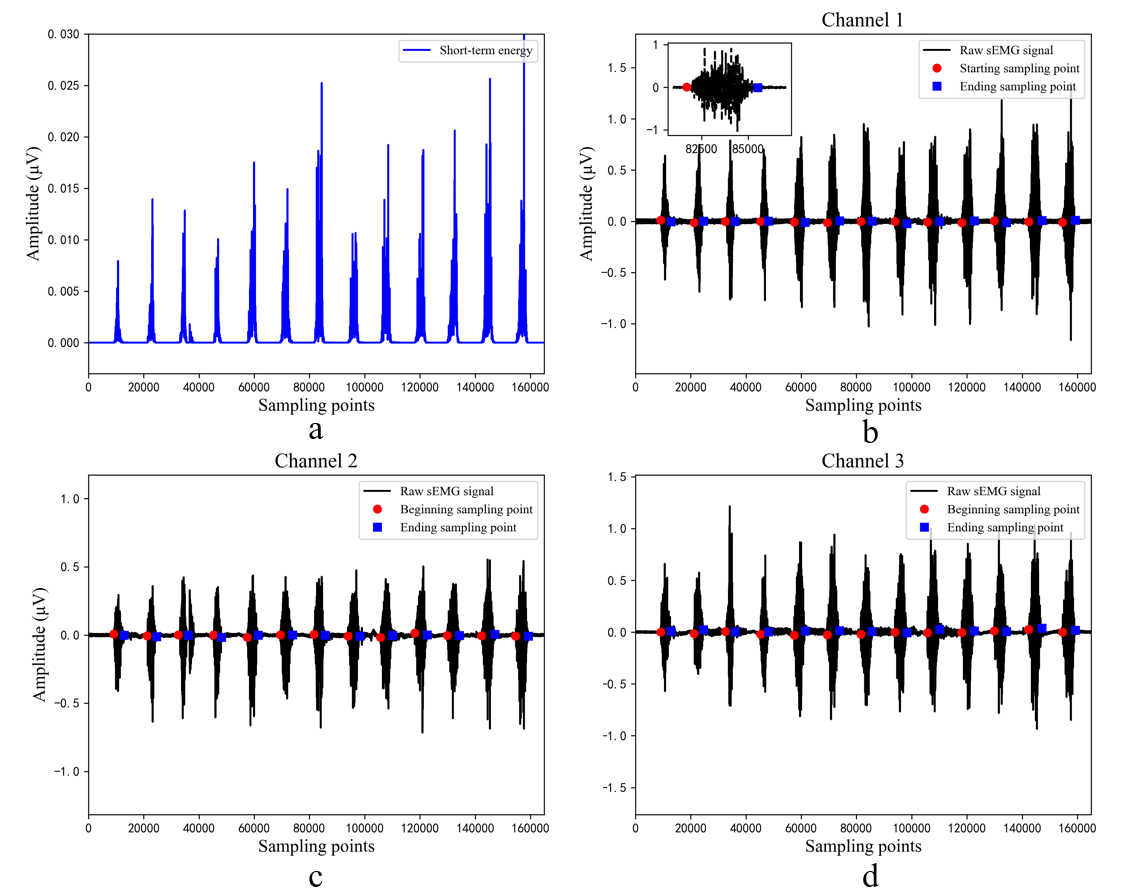}
	\caption{(a) Distribution of short-time energy; (b) Channel 1 for active segment detection; (c) Channel 2 for active segment detection; (d) Channel 3 for active segment detection}
	\label{fig-6}
\end{figure}

Step 3: Calculate the data point difference L between the action start point $x_{s}$ and the action endpoint $x_{e}$:
\begin{equation}
	L=e-s
	\label{eq：2}
\end{equation}
Considering that the duration of the sEMG signal for each leg action is between 1000 ms and 2000 ms, if L $>$ 2000 and L $<$ 4000, then $x_{s}$ is the true action start point and $x_{e}$ is the true action end point. Otherwise, we consider the data in the segment to be fluctuation noise. Fig \ref{fig-6} shows the process of calculating short-term energy and extracting active segments from the three channels.
\subsection{Feature extraction}

In this paper, features were extracted from channel 1 and channel 2 of the sEMG signal, a total of 44 features. Each channel has 22 features, consisting of 17 time-domain features and 5 frequency-domain features.

The time domain analysis is to consider the sEMG signal as a time-dependent signal and computes time statistics based on amplitude values \cite{jose2017classification}. Frequency domain analysis is the Fourier transform of the sEMG signal to analyze its frequency characteristics.

\subsection{Methods}
\subsubsection{Supervised learning}

In order to obtain the best lower limb movement recognition rate, we investigated five supervised learning algorithms (classifiers) including BPNN, CNN, LDA, LSTM, and SVM\cite{toledo2019support,zhang2013adaptation,chaobankoh2022lower,wang2020semg,ahsan2011neural}.

In this paper, we first perform normalization preprocessing on the input feature data. This preprocessing can not only solve the problems caused by the different magnitudes of the parameters, but also help to optimize the performance of the neural network. Through normalization preprocessing, the range of feature data is scaled to between [0,1] with the following formula:
\begin{equation}
	{{\bar{x}}_{i}}=\frac{{{x}_{i}}-{{x}_{\min }}}{{{x}_{\max }}-{{x}_{\min }}}
	\label{eq：3}
\end{equation}

Where $x_{i}$ denotes the input data, $x_{max}$ denotes the maximum input and $x_{min}$ denotes the minimum input.
\subsubsection{SVM-RFE feature selection}

This study uses the SVM-RFE algorithm for feature selection to select the optimal subset of features. The SVM-RFE feature selection algorithm is a backward sequential approximation and reduction algorithm based on the principle of maximizing the intervals in the SVM \cite{zhang2009feature}. The basic steps are: First, train the SVM model using the feature set to obtain the initial SVM model parameters; second, select an appropriate ranking criterion and compute the ranking criterion scores for all features in the current feature set; and finally, remove the features with the lowest scores from the feature set. The above three steps are performed cyclically until the optimal feature set is filtered. In the process of calculating the feature ranking criterion scores, the cost function as in Eq. \ref{eq：4} is often chosen as the ranking criterion.
\begin{equation}
	{{R}_{C}}=\left| {{S}^{2}}-{{S}^{-{{(P)}^{2}}}} \right|
	\label{eq：4}
\end{equation}

In Eq. \ref{eq：4},$ {{S}^{2}}$ represents the initial weights of the SVM model, while  representing the weights of the SVM model obtained after removing the P-th feature, and the expression is as follows:
\begin{equation}
	{{S}^{2}}=\sum\limits_{i,j=1}^{N}{\alpha _{i}^{*}\alpha _{j}^{*}{{y}_{i}}{{y}_{j}}K\left( {{x}_{i}},{{x}_{j}} \right)}
	\label{eq：5}
\end{equation}

In Eq. \ref{eq：5}, N denotes the number of samples, i and j denote cyclic variables, y denotes the action category, and $K\left( {{x}_{i}},{{x}_{j}} \right)$ denotes the kernel function for SVM classification,  ${\alpha _{i}^{*}}$ and ${\alpha _{j}^{*}}$  can be obtained by solving the SVM pairwise problem.

\begin{figure}[htbp]
	\centering
	\includegraphics[scale=0.8]{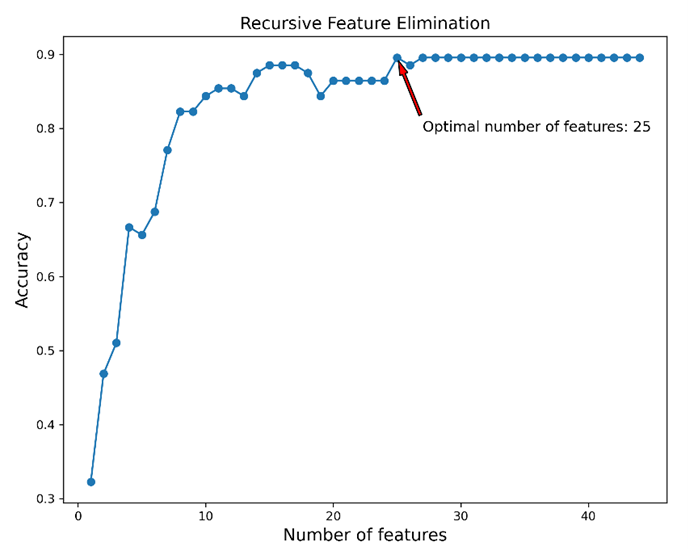}
	\caption{SVM-RFE selects the optimal number of features}
	\label{fig-7}
\end{figure}

In this paper, sample data from five subjects, comprising a total of 2400 samples, was used to construct the feature set. Finally, from 44 features, the top 25 features with the highest importance were selected to form the optimal feature subset for subsequent classification tasks. The selection of the optimal number of features is shown in Fig \ref{fig-7}.
\section{Results}

\begin{figure}[htbp]
	\centering
	\includegraphics[scale=0.8]{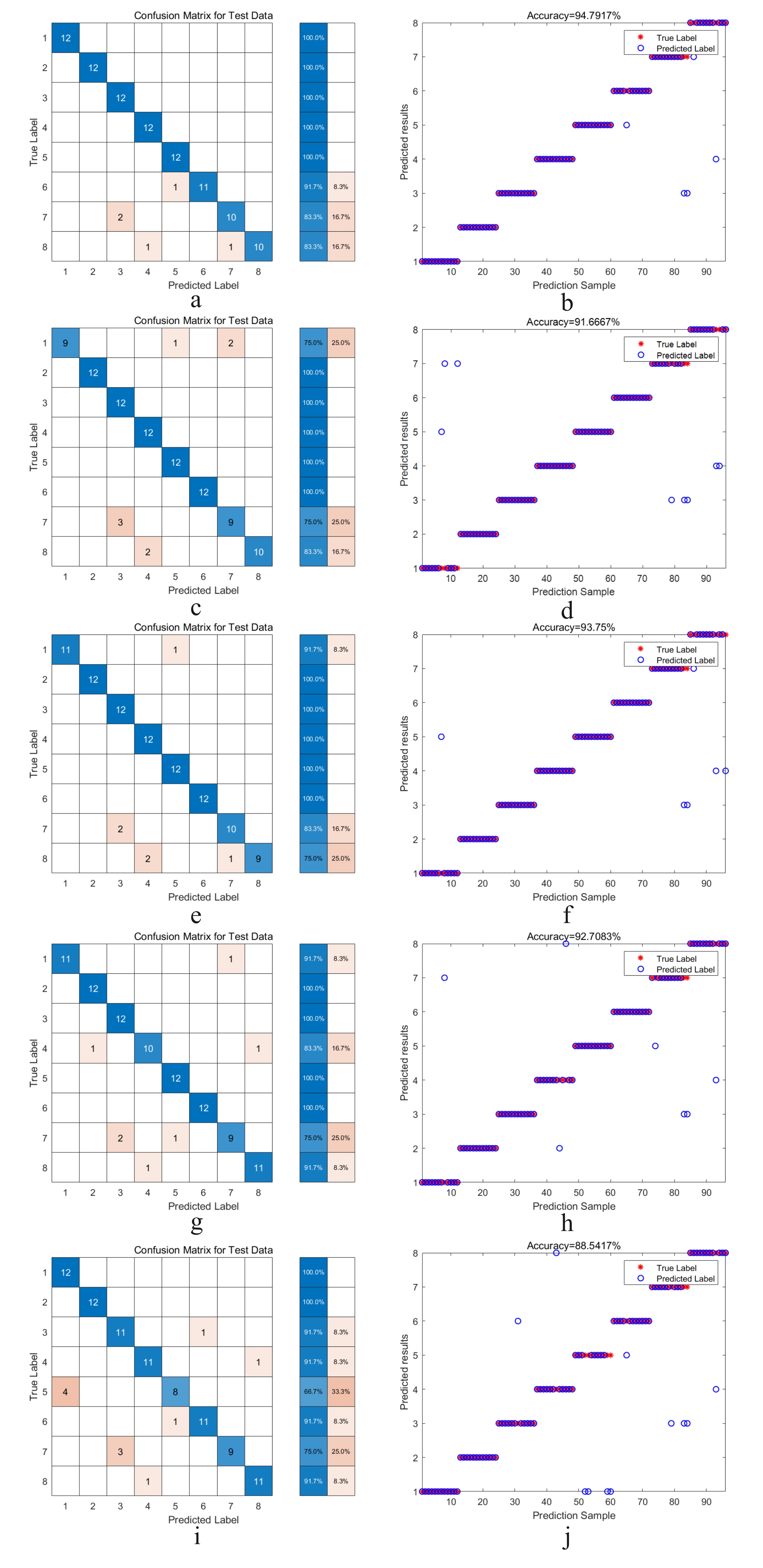}
	\caption{5 Supervised learning classification confusion matrices with classification results. }
	\label{fig-9}
\end{figure}

In this study, eight different lower limb movements were classified using five supervised learning classifiers. The classification results of the eight lower limb movements of one of the subjects are shown in Fig \ref{fig-9}.Fig \ref{fig-9}(a) illustrates the action classification confusion matrix of BPNN,Fig \ref{fig-9}(b) illustrates the classification results and average recognition rate of the BPNN model with an average action recognition rate of 94.79\%; Fig \ref{fig-9}(c) represents the action classification confusion matrix of CNN, Fig \ref{fig-9}(d) illustrates the classification results and average recognition rate of the CNN classifier with an average action recognition rate of 91.67\%;Fig \ref{fig-9}(e) represents the action classification confusion matrix of LDA, while Fig  \ref{fig-9}(f) illustrates the classification results and average recognition rate of LDA classifier, with an average action recognition rate of 93.75\%;Fig \ref{fig-9}(g) represents the action classification confusion matrix of the LSTM, and Fig \ref{fig-9}(h) illustrates the classification results and average recognition rate of the LSTM classifier with an average action recognition rate of 92.71\%;Fig \ref{fig-9}(i) illustrates the action classification confusion matrix of SVM, and Fig \ref{fig-9}(j) illustrates the classification results and average recognition rate of SVM classifier, with an average action recognition rate of 88.54\%. Fig \ref{fig-9}shows that within the classification model, these models may misclassify action A7 as action A3 and action A8 as action A4. This error may be related to the similarity in the position of the muscles. In particular, in the SVM classifier, action A5 is also misclassified as action A1, which leads to a decrease in the average recognition rate of action.
\begin{table}[h!]
	\centering
	\caption{Accuracy of six classifiers}
	\begin{tabular}{cccccc}
		\hline
		\textbf{Subject/Classifier} & \textbf{BPNN} & \textbf{CNN} & \textbf{LDA} & \textbf{LSTM} & \textbf{SVM} \\ \hline
		\textbf{1}                  & 94.79         & 91.67        & 93.75        & 92.71         & 88.54        \\
		\textbf{2}                  & 92.71         & 85.42        & 92.71        & 87.5          & 89.58        \\
		\textbf{3}                  & 92.71         & 90.63        & 92.05        & 92.71         & 91.67        \\
		\textbf{4}                  & 96.88         & 92.71        & 93.75        & 91.67         & 88.54        \\
		\textbf{5}                  & 97.92         & 96.88        & 93.75        & 94.79         & 97.92        \\
		\textbf{Average (\%)}       & 95.00         & 91.46        & 93.20        & 91.88         & 91.25        \\ \hline
	\end{tabular}
	\label{tab4}
\end{table}

\begin{figure}[!t]
	\centering
	\includegraphics[scale=0.55]{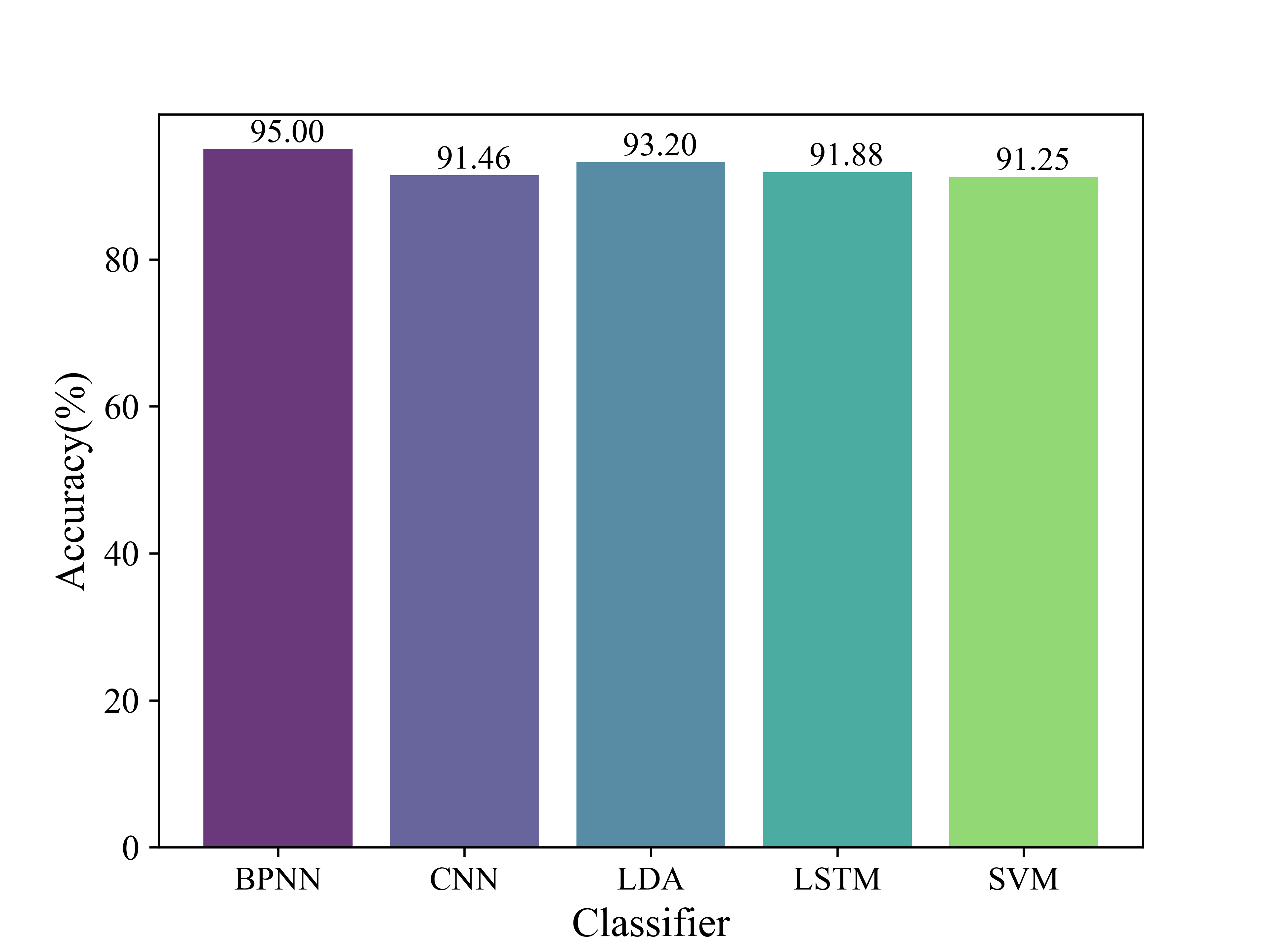}
	\caption{Average recognition rate of 5 classifiers}
	\label{fig-10}
\end{figure}

Table \ref{tab4} shows the average recognition performance of the eight lower limb movements for the five subjects using five different supervised learning methods. From the data in Fig \ref{fig-10}, it is clear that the average recognition rate of BPNN is as high as 95\%, while the average recognition rate of CNN is 91.46\%, LDA is 93.20\%, LSTM is 91.88\% and SVM is 91.25\%. Compared to the other four supervised classifiers, the experimental results show that BPNN has the best recognition results. This advantage is attributed to the superior non-linear modeling capability of BPNN, which helps to capture complex data patterns; the multi-layer structure, which enables multi-level data representation; and the ability to continuously optimize the model through the back-propagation algorithm. In summary, BPNN is more advantageous than the other four algorithms in this paper in terms of recognition effectiveness and demonstrates a more powerful generalization capability.

\section{Conclusions and discussion}

In this study, we innovatively combine SVM-RFE with BPNN for lower limb movements recognition based on the sEMG signal. The features extracted from the sEMG signal are filtered by the SVM-RFE method, and then the eight lower limb movements are detected by the BPNN classifier. The results of the study proved that the average recognition rate using the combination of SVM-RFE and BPNN is as high as 95\%. Compared to existing methods, this study has obvious advantages in terms of muscle selection, sensor type, feature selection algorithm, and motion recognition accuracy.

This study is expected to improve the quality of life of patients with lower limb movement disorders and lower limb amputees, and provide a reference for the design and research of intelligent wearable lower limb prostheses. However, this study has some limitations. First, this study only considered specific lower limb movement contexts and therefore is not applicable to complex lower limb movement contexts. Real-life activities such as walking, running, climbing, and various other activities require further research and testing. Second, the study had a limited number of participants and did not consider the participation of people with disabilities. To ensure the applicability of this study, data from people with different levels of disability will be added later. Finally, it is worth noting that the data processing and analysis in this study was limited to offline situations and has not yet been validated in an online system, which is a direction that could be explored in future studies to more fully assess the validity of the study.

\bibliographystyle{IEEEtran}
\bibliography{IEEEabrv, mybibfile.bib}

\end{document}